# Introducing DEFORMISE: A deep learning framework for dementia diagnosis in the elderly using optimized MRI slice selection

Nikolaos Ntampakis [a,b,*,1], Konstantinos Diamantaras [a], Ioanna Chouvarda [c], Vasileios Argyriou [d], Panagiotis Sarigiannidis [b,e]

[a] International Hellenic University, Sindos, Greece
[b] MetaMind Innovations, Kozani, Greece
[c] Aristotle University of Thessaloniki, Thessaloniki, Greece
[d] Kingston University London, London, UK
[e] University of Western Macedonia, Kozani, Greece

ABSTRACT

Dementia, a debilitating neurological condition affecting millions worldwide, presents significant diagnostic challenges. In this work, we introduce DEFORMISE, a novel DEep learning Framework for dementia diagnOsis of eldeRly patients using 3D brain Magnetic resonance Imaging (MRI) scans with Optimized Slice sElection. Our approach features a unique technique for selectively processing MRI slices, focusing on the most relevant brain regions and excluding less informative sections. This methodology is complemented by a confidence-based classification committee composed of three novel deep learning models. Tested on the Open OASIS datasets, our method achieved an impressive accuracy of 94.12%, surpassing existing methodologies. Furthermore, validation on the ADNI dataset confirmed the robustness and generalizability of our approach. The use of explainable AI (XAI) techniques and comprehensive ablation studies further substantiate the effectiveness of our techniques, providing insights into the decision-making process and the importance of our methodology. This research offers a significant advancement in dementia diagnosis, providing a highly accurate and efficient tool for clinical applications.

## 1. Introduction

Dementia, a progressive neurological disorder, significantly impairs cognitive function, affecting millions of individuals globally. It poses a substantial burden on patients, caregivers and healthcare systems. Alzheimer's disease, the most common form of dementia, accounts for 60%–70% of cases [1]. The strongest known risk factor for dementia is increasing age, with most cases affecting those of 65 years and older [2]. Early and accurate diagnosis of dementia is crucial for effective management and care planning [3].

According to Smith et al. [4] the advancements in neuroimaging, particularly Magnetic Resonance Imaging (MRI), have revolutionized the diagnostic landscape for dementia. MRI provides detailed brain images, facilitating the identification of structural changes associated with various dementia stages. However, the interpretation of these images typically relies on the expertise of radiologists and the accuracy can vary depending on their level of experience.

Medical imaging diagnostic efficiency and accuracy may be improved by recent advances in machine learning [5]. Machine learning algorithms, particularly those requiring classification problems, have been successfully used to identify between patients who are demented and those who are not, according to Chiu et al. [6]. Despite these developments, current approaches frequently make use of whole MRI scans, which results in the addition of extraneous data that may compromise the precision and effectiveness of these algorithms.

The decision to classify patients into two broad categories – demented and non-demented – is primarily to robustly determine the presence or absence of dementia in a clinical setting. This binary classification lays the groundwork for clear, actionable diagnostic outcomes. Once dementia is confirmed, we can then focus on the more nuanced task of determining the stage and type of dementia. A study by Tufail et al. [7] highlights the effectiveness of binary classification in Alzheimer's disease diagnosis using structural MRI and deep learning. They note that conventional methods, which often require expert interpretation and feature extraction, can be enhanced by machine learning approaches that effectively distinguish between Alzheimer's disease and healthy subjects. Their findings underscore the advantages of binary classification in terms of simplicity and efficiency, particularly when leveraging advanced machine learning techniques for diagnostic purposes.

* Corresponding author.
*E-mail addresses:* nikontam1@iee.ihu.gr, nntampakis@metamind.gr (N. Ntampakis).
[1] Corresponding addresses: International Hellenic University, Sindos, Greece & MetaMind Innovations, Kozani, Greece.

Among the most common datasets used for dementia-related machine learning research are the Open Access Series of Imaging Studies (OASIS) and the Alzheimer's Disease Neuroimaging Initiative (ADNI). OASIS-1 is a cross-sectional dataset [8], OASIS-2 includes longitudinal data [9] and ADNI is a study that has generated a dataset widely used in dementia-related research [10].

In response to these challenges, this paper introduces DEFORMISE. This approach leverages the strengths of 3 models, enhancing overall accuracy through collective decision-making. A significant innovation of our work is the selective processing of MRI scans, focusing exclusively on slices that predominantly feature the brain and excluding irrelevant sections. This targeted approach, validated on the OASIS dataset, not only improves diagnostic accuracy, but also enhances computational efficiency.

The contributions of this paper can be summarized as follows:

- Our selective processing of MRI scans excludes less informative sections, thereby reducing computational load and improving diagnostic precision.
- The methodology has been tested on the Open Access Series of Imaging Studies (OASIS) datasets and validated in Alzheimer's Disease Neuroimaging Initiative (ADNI) dataset. This significant improvement not only sets a new benchmark in the domain but also offers a more efficient pathway for clinical implementations.
- Our selective processing approach can be applied to all MRI scans conducted using the same protocol, specifically the MPRAGE (Magnetization - Prepared Rapid Gradient Echo) protocol. This adaptability enhances the potential for broader clinical application and standardization in MRI analysis.

The rest of the paper is organized as follows: Section 2 presents the Related Work. Section 3 describes the dataset used, providing details on the data used in our study. Section 3 discusses the Methodology, explaining the approaches and techniques employed. Section 4 presents the Experimental Outcomes including an Ablation Study and Section 5 concludes the paper with a summary of our work and potential avenues for future research.

## 2. Related work

The body of the research about the related work is focusing on the binary classification (demented vs. non-demented) of 3D brain MRIs, particularly using the OASIS dataset. Dhinagar et al. [11] conducted a study that highlighted the capability of a 3D CNN model trained from scratch. This model achieved an ROC-AUC of 0.789 for Alzheimer's Disease (AD) classification, demonstrating its ability to generalize across datasets, including handling diverse MRI data from the OASIS dataset. The model showed less susceptibility to overfitting in AD classification compared to Parkinson's Disease (PD) and proposed that such a model could be instrumental in differentiating between AD and PD, especially in complex cases where both diseases present similar symptoms.

In another significant work, Yagis et al. [12] focused on the early detection of AD using deep learning and neuroimaging data. Their study emphasized the crucial role of early AD detection, given that diagnostic symptoms often emerge at later stages after substantial neural damage. They employed a 3D VGG variant convolutional network (CNN) for analyzing MRI data, underlining the potential of deep learning in extracting high-level features from neuroimaging. Using the OASIS dataset, they aimed to improve the accuracy of AD classification. The researchers preferred 3D models over 2D to prevent the loss of information, which is common when converting 3D MRIs into 2D images for analysis. Their proposed 3D CNN model achieved a classification accuracy of 69.9% on the OASIS dataset, using 5-fold cross-validation, which was notably superior to that of 2D network models.

Saratxaga et al. [13] presented a study that developed a deep learning-based method for AD diagnosis using the OASIS neuroimaging dataset. Their experiments tested various models on both OASIS-1 and OASIS-2 datasets for two-class (cognitive normal vs. AD) and three-class (cognitive normal vs. very mild vs. mild and moderate dementia) problems. Notably, the study found that 2D network models in binary classification problem, particularly BrainNet2D and ResNet18, showed accuracy of 0.83 and 0.93 respectively, surpassing previous approaches. At the same time BrainNet3D achieved an accuracy of 0.84. The study concluded that the 2D approach was more efficient for both binary and multi-class classifications.

Furthermore, Wen et al. [14] developed an open-source framework for the classification of Alzheimer's disease using 3D convolutional neural networks. This framework extended existing tools to include data from ADNI, AIBL, and OASIS datasets, and focused on 3D subject-level analysis for AD classification. Testing on the OASIS dataset showed an accuracy of 0.68, highlighting the framework's effectiveness in AD classification using deep learning techniques.

It is noteworthy that there have been important results using 2D data in the realm of Alzheimer's disease diagnosis and classification using brain MRIs. Maqsood et al. [15] developed an automated detection and classification system for the early diagnosis of Alzheimer's disease through brain MRI. The study leveraged transfer learning, fine-tuning the pre-trained AlexNet convolutional network for image classification. The system was designed to classify dementia patients and identify the four stages of dementia progression, including binary and multi-class classification tasks, using both segmented and un-segmented brain MRI data. The results of the system using the OASIS dataset were encouraging. Its total accuracy for multi-class classification of unsegmented pictures was 92.85%. The system achieved an accuracy of 89.66% in binary classification.

Building on the discussions around studies using the OASIS dataset for Alzheimer's disease classification, it is important to note that other studies in the field have utilized different datasets, such as the ADNI dataset [10]. ADNI dataset is a multi-site study initiated in 2004, focuses on tracking Alzheimer's disease progression using neuroimaging, biochemical, and genetic markers. It includes data from subjects with AD, those who may develop AD, and controls with no signs of cognitive impairment. While this dataset is closely align with research in dementia at large, its primary emphasis is on Alzheimer's disease, a subset of dementia. Dementia, as a broader category, encompasses various forms that may not necessarily be Alzheimer's. One of the most promising is the study conducted by Ebrahimi et al. [16] With an emphasis on 3D classification and the usage of the ADNI dataset, Ebrahimi investigated the efficacy of convolutional neural networks (CNNs) in detecting Alzheimer's disease using magnetic resonance imaging. Their study compared multiple models, such as recurrent neural networks (RNNs), three-dimensional (3D) CNNs, and two-dimensional (2D) CNNs. The study brought transfer learning from 2D pictures to improve the performance of 3D CNNs and emphasized the shortcomings of 2D CNNs in processing 3D MRI volumes. Voxel-based judgments were made possible by this method, which improved classification accuracy. The astounding outcomes demonstrated how much better the 3D CNN with transfer learning performed than the other approaches. When used on the ADNI dataset, it demonstrated an 96.88% accuracy, 100% sensitivity, and 94.12% specificity in differentiating between AD patients and healthy people.

It is essential to mention our approach towards the selection of studies for inclusion in our analysis. Notably, we had to exclude certain papers from the review of related work due to concerns over data leakage, particularly in studies dealing with 2D images derived from 3D volumetric data. Data leakage, a critical issue in machine learning, refers to the inadvertent inclusion of data in both the training and testing datasets, which can lead to overly optimistic estimates of a model's performance. In the context of our study, we identified instances where 2D images, extracted from the same 3D volumetric scans of a patient, could be presented in training, validation, and testing datasets simultaneously. This overlap can result in models that appear highly accurate but are, in fact, simply recognizing repeated data. Moreover, we also chose to exclude studies where data was sourced from non-trusted or unofficial sources.

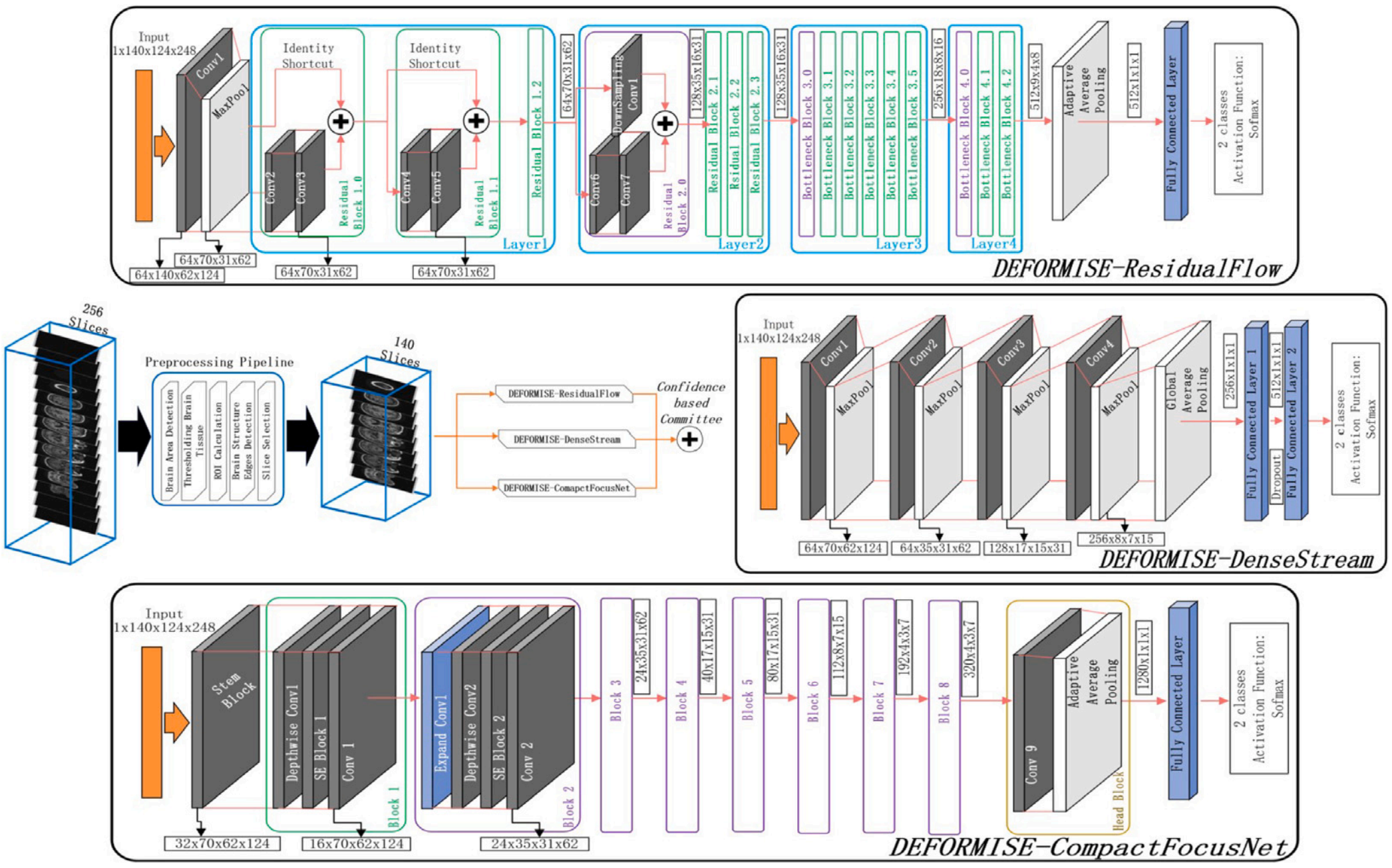

**Fig. 1.** DEFORMISE methodology.

## 3. Methodology

In our study, as shown in Fig. 1, we employed a comprehensive methodology that starts with the processing of volumetric brain MRI scans of patients. These MRIs consist of continuous slices, providing an intricate view of the brain's structure. Our aim was to develop a model capable of distinguishing between demented and non-demented patients based on these scans.

In the first phase of our methodology we aimed to maintain a homogeneous dataset, as OASIS2 includes patients above 60 years old only when all patients below 60 were non-demented in the OASIS1 dataset, avoiding in this way to introduce age-related biases. Then, we implemented a novel technique to select the 140 most relevant continuous slices from each brain MRI. Concurrently, we performed minor transformations on the data.

Following the preprocessing, the manipulated volumetric data were passed onto a confidence-based committee. This committee utilized three distinct custom 3D deep learning models: a custom 3D Convolutional Neural Network (CNN) called "DEFORMISE-DenseStream", a custom 3D CNN based on residual connections called "DEFORMISE-ResidualFlow" and a custom 3D CNN based on compact bottleneck structures called "DEFORMISE-CompactFocusNet". Each model generated predictions with associated confidence levels. The committee evaluated the predictions from each model, taking into account the confidence associated with each. Based on this analysis, a final decision was made to classify each patient as either demented or non-demented.

### 3.1. Preprocessing pipeline

#### 3.1.1. Data cleaning

During the preprocessing pipeline for this research, a selection process was undertaken to ensure data consistency and avoid data leakage, when combining OASIS1 and OASIS2 datasets.

For the OASIS1 dataset, which initially comprised 416 patients, the selection criteria focused on including only those patients with 'MR1' scans (excluding 'MR2' scans[2]) and specifically choosing from the 'RAW' files only the first session ('mpr-1'). This filtering was applied to prevent data leakage by avoiding the inclusion of multiple sessions per patient, which could lead to overlapping information in the dataset. As a result of this filtering, all 416 patients from the OASIS1 dataset were retained.

In the case of the OASIS2 dataset, which contains longitudinal data for 150 patients, a similar approach was adopted. To maintain consistency and prevent data leakage, only the first visit of each patient was selected, applying the same criteria as used for OASIS1. This procedure resulted in the inclusion of 146 patients from the OASIS2 dataset. Four patients were excluded because they did not have the ('mpr-1') session in their raw data.

An additional filtering step was applied to the OASIS1 dataset, considering the age range of the subjects. Since OASIS1 includes patients from ages 18 to over 90, and OASIS2 is focused on subjects aged 60 and above, it was decided to align both datasets by age range. Consequently, patients under the age of 60 were removed from the OASIS1 dataset, reducing the number of patients from 416 to 198. This adjustment ensured consistency with the age range of the OASIS2 dataset and facilitated a more focused study on the older age group, which is more relevant for dementia research.

After these selection processes, the combined dataset resulted in a total of 344 unique patients (OASIS1 and OASIS2 contain non-overlapping cohorts by design, ensuring no patient appears in both datasets). The next step was to classify these patients into two groups: demented and non-demented, based on their CDR. Patients with a CDR of 0.5, 1, and 2 were categorized as demented, while those with a CDR of 0 were considered non-demented. This classification yielded 164 demented and 180 non-demented patients (total 354 patients), providing a balanced dataset for further analysis in the study.

[2] 'MR2' refers to reliability scans taken from 20 of the initial subjects. These scans aim to benchmark the reliability of analytic procedures, with differences in images largely attributable to factors like head positioning or scanner variability.

#### 3.1.2. Extraction

The next step involved extracting slices from the NIfTI files for each patient. For this task, the 'nibabel' library, a widely recognized toolkit for NIfTI data handling in Python, was utilized to process the MRI data [17]. Using this tool, all the available 256 slices from the NIfTI file of dimensions 248 × 496 pixels were extracted from each patient's MRI scans and converted into a 2D format.

#### 3.1.3. Optimized subset selection

At the core of the preprocessing pipeline in this study lies the selection of a subset of 256 slices from the volumetric MRI data, chosen for their high predictive value. This selection is critical as, based on MRI observations, the topmost and bottom-most slices often contain minimal relevant information, contributing little to the predictive model's accuracy. Researches by Lee et al. [18] and Im et al. [19] suggest that the average human head height is around 17.5 cm. Since the MPRAGE protocol produces 1.25 mm thick slices, covering 17.5 cm requires exactly 140 slices (175 mm ÷ 1.25 mm = 140). This determines the optimal slice count based on anatomical constraints, ensuring complete brain coverage from foramen magnum to vertex. The full MRI volume contains 256 slices, but many capture non-brain regions. Therefore, we need to identify which continuous 140-slice window within the 256 slices contains the most diagnostic information. This focused selection not only ensures the inclusion of the brain's most relevant regions but also enhances the efficiency of training the predictive models. By reducing the dataset to slices that are most likely to contain significant features for dementia detection, the training process becomes more streamlined and focused.

$$\sigma_B^2(t) = \omega_0(t)\omega_1(t)[\mu_0(t) - \mu_1(t)]^2 \tag{1}$$

In order to apply this optimal subset selection, the initial procedure focuses on detecting the Region of Interest (ROI). This ROI will be the sub-area of our calculations. To do that, we applied Otsu's method [20]. As shown in (1), Otsu's method thresholds the images to create a binary distinction between the brain and the background, crucial for accurate region identification. Otsu's method determines an optimal threshold for segmenting images into brain and background areas. This is achieved by maximizing the between-class variance, denoted as $\sigma_B^2(t)$:

This variance is a measure of separation between two classes, defined by their pixel intensity distributions. The probabilities of these two classes, given by $\omega_0(t)$ for the background and $\omega_1(t)$ for the brain, are calculated based on the threshold $t$. The mean intensities of the two classes are represented by $\mu_0(t)$ and $\mu_1(t)$. By optimizing the threshold to maximize $\sigma_B^2(t)$, Otsu's method transforms the gray-scale image into a binary one.

Next step is to calculate the bounding box for each slice of the same patient. The focus here is to select the largest bounding box from these slices as the ROI for that particular patient. In this way we ensure that our ROI will be shorter from the whole image and will include all the MRIs essential information. After applying Otsu's method for thresholding the MRI images, we label the pixels and we calculate the bounding box, which is the smallest rectangle that can enclose the labeled region. The margins of this bounding box are determined by the extremities of the region. Fig. 2 shows an indicative ROI calculated out of the slices of a patient.

The coordinates for the bounding box are calculated based on (2) using the minimum and maximum x and y coordinates of the pixels within the region, where $\min_x$ and $\min_y$ represent the minimum x and y coordinates of all the pixels, forming the top-left corner of the bounding box. Similarly, $\max_x$ and $\max_y$ are the maximum x and y coordinates, representing the bottom-right corner of the bounding box.

$$\begin{aligned} \min_x &= \min(x_i \mid (x_i, y_i) \in \text{Region}) \\ \min_y &= \min(y_i \mid (x_i, y_i) \in \text{Region}) \\ \max_x &= \max(x_i \mid (x_i, y_i) \in \text{Region}) \\ \max_y &= \max(y_i \mid (x_i, y_i) \in \text{Region}) \end{aligned} \tag{2}$$

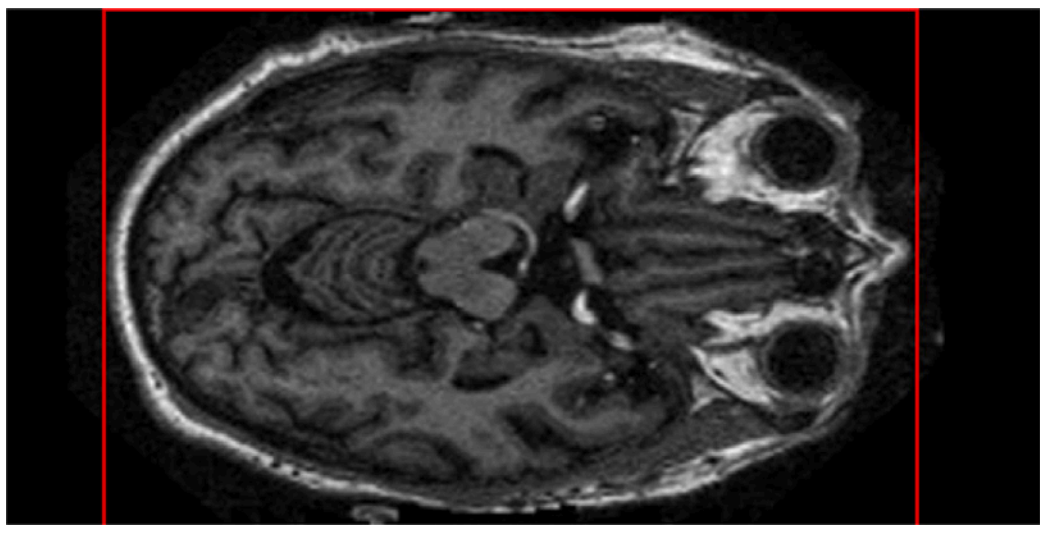

**Fig. 2.** Indicative ROI.

Canny Edge Detection, is employed as the next step, as it is pivotal for quantifying the edges within the ROI in patient's MRI slices [21]. This edge detection algorithm excels in detecting sharp changes in image intensity, indicative of object boundaries. The Canny Edge Detection algorithm works through a series of steps, each contributing to the identification of edges in an image. Initially as shown in (3) the process starts by applying a Gaussian filter to the image to reduce noise.

$$G(x, y) = \frac{1}{2\pi\sigma^2} e^{-\frac{x^2+y^2}{2\sigma^2}} \tag{3}$$

where $x$ and $y$ are the distances from the origin in the horizontal and vertical axes, respectively, and $\sigma$ is the standard deviation of the Gaussian distribution. Then, following (4) and (5), the calculation of the gradient of the image intensity is taking place where * denotes convolution and $A$ is the image.

$$G_x = \begin{bmatrix} -1 & 0 & +1 \\ -2 & 0 & +2 \\ -1 & 0 & +1 \end{bmatrix} * A \tag{4}$$

$$G_y = \begin{bmatrix} -1 & -2 & -1 \\ 0 & 0 & 0 \\ +1 & +2 & +1 \end{bmatrix} * A \tag{5}$$

Next, a non-maximum suppression is taking place for thinning out the edges. It checks each pixel in the gradient image and retains it only if it is a local maximum in the direction of the gradient. Finally, the algorithm uses two thresholds to differentiate between strong and weak edges in line with common practices in image analysis. Strong edges are marked where the gradient magnitude exceeds the high threshold, specifically set at 20% of the maximum image intensity and the weak edges are identified where the gradient magnitude is between the high and low thresholds, with the low threshold set at 10% of the maximum intensity. These thresholds are standards in edge detection, providing a balance between capturing essential features and minimizing noise. Weak edges are only retained if they are connected to strong edges. The choice of Canny Edge Detection is strategic, as it effectively highlights structural features in MRI slices, such as the brain's boundaries.

Thus, we calculate the sum of Canny edges within the ROI of each slice. This sum serves as a numerical representation of edge density or detail level in each slice. The distribution of these values across continues slices of the same patient follows a pattern akin to a Gaussian distribution, Fig. 3. The edge sum quantifies anatomical complexity - slices with higher edge density contain more structural boundaries (gray-white matter interfaces, ventricular edges, cortical folding) that are diagnostically relevant. We select the 140 continuous slices that maximize the cumulative edge sum, ensuring capture of the most anatomically complex brain regions.

The lower slices, typically showing minimal brain tissue, exhibit fewer edges. As the slices progress to the middle of the brain, where structures like the nose, eyes, and brain tissues are more pronounced, the edge sum reaches its peak, indicating the highest complexity and detail level. Beyond this region, towards the top of the head, the complexity decreases again, mirrored by a reduction in the sum of edges.

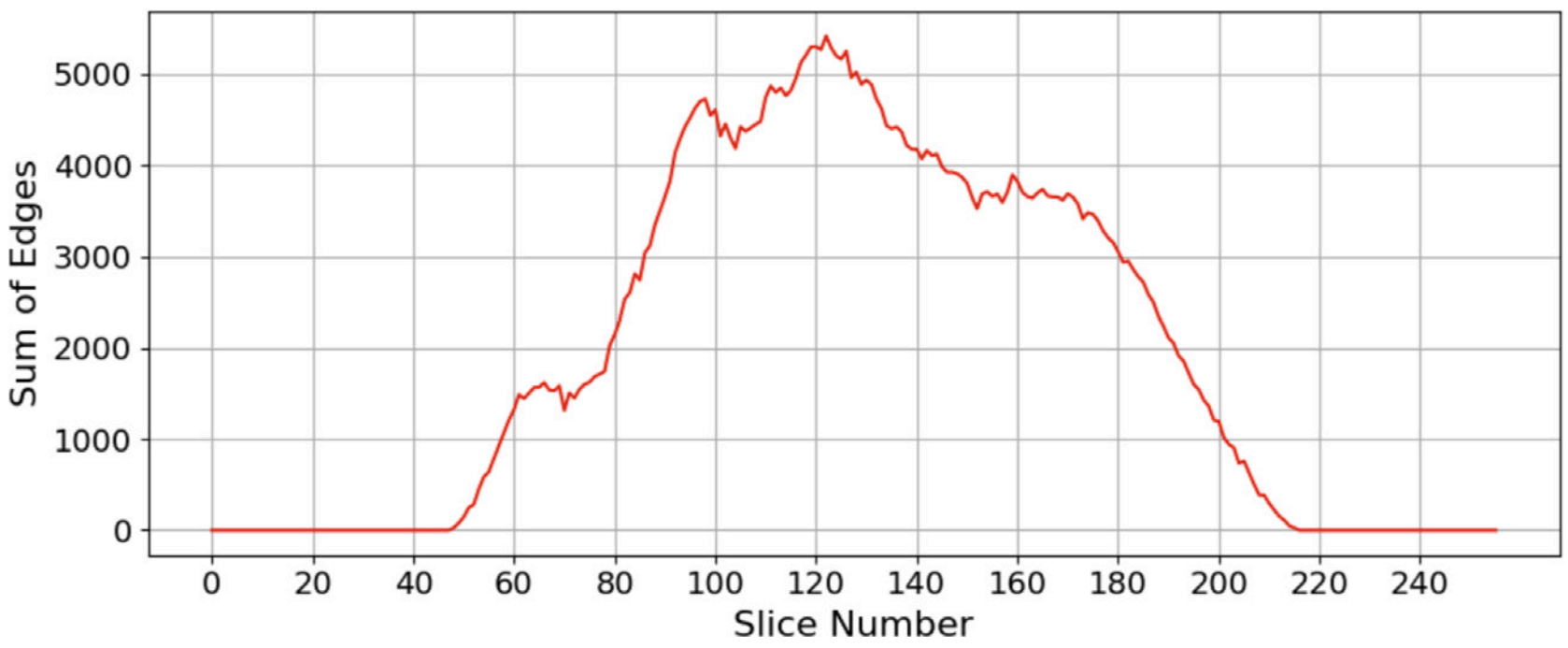

**Fig. 3.** Indicative Canny Edges sum per slice for 0124 patient of OASIS1 dataset.

Based on the analysis incorporating Otsu's method and Canny Edge Detection, 140 continuous slices with maximal information were selected, as shown in Fig. 1. Post-hoc manual validation confirmed that our edge-based selection consistently captured all critical structures: across 344 patients, 100% of selected windows included the complete hippocampus and entorhinal cortex, while excluded slices contained primarily skull, neck tissue, or empty space. These slices, identified for their significant Canny Edge sums, ensure the inclusion of the most informative brain regions.

#### *3.1.4. Transformation*

The MRI images were resized from $248 \times 496$ to $124 \times 248$ pixels, resulting in volumetric data per patient of $140 \times 124 \times 248$. This anisotropic downsampling (maintaining higher resolution along the slice axis at 1.25 mm while reducing in-plane resolution) was adopted to balance computational memory constraints with preservation of critical anatomical information. Isotropic downsampling to 1.25 mm resolution would require $140 \times 248 \times 496$ voxels, necessitating approximately 4× more GPU memory. Our architectural adaptations, particularly the asymmetric strides (1,2,2) in residual convolution layers, explicitly account for this data geometry. For the train-test split, a stratified approach was employed with a 90/10 ratio, leading to 310 patients in the training set and 34 in the test set.

### *3.2. Confidence-based ensemble prediction method*

#### *3.2.1. DEFORMISE-ResidualFlow*

The DEFORMISE-ResidualFlow model architecture was inspired by ResNet [22] principles but represents a custom 3D implementation specifically designed for volumetric MRI analysis, with architectural details visualized in Fig. 1. Unlike standard ResNet which uses uniform stride patterns, our model employs asymmetric strides in the initial convolution to account for the anisotropic nature of our preprocessed data ($140 \times 124 \times 248$ voxels). The core component is a 3D CNN, highly effective for visual data processing [23]. The initial convolution layer uses a $7 \times 7 \times 7$ kernel with a stride of $(1, 2, 2)$, moving across depth, height, and width. Padding is set to 3 to handle input edges. After this, Batch Normalization (BN) [24] is applied, as shown in Eq. (6), followed by the ReLU activation function in Eq. (7), and then a 3D max pooling layer with kernel size $3 \times 3 \times 3$, stride 2, and padding 1.

$$\mathrm{BN}(x) = \gamma \left( \frac{x - \mu}{\sqrt{\sigma^2 + \epsilon}} \right) + \beta \tag{6}$$

$$\mathrm{ReLU}(x) = \max(0, x) \tag{7}$$

The core of the network comprises 4 residual layers. The first has 3 blocks, each with two 3D convolutional layers ($3 \times 3 \times 3$), followed by batch normalization and ReLU activation. Residual connections add the input directly to the output of the convolutional layers, maintaining spatial dimensions, as shown in Eq. (8).

$$\mathrm{y} = F(\mathrm{x}, W_i) + \mathrm{x} \tag{8}$$

The second residual layer consists of 4 blocks with 128 filters and down-samples spatial dimensions using a $1 \times 1 \times 1$ convolution. The third layer adds 6 blocks with 256 filters, further down-sampling. The fourth layer includes 3 blocks with 512 filters, continuing the down-sampling process. Residual connections ensure consistency in feature map dimensions. After the residual layers, an Adaptive Average Pooling layer adjusts the feature maps to a fixed size $1 \times 1 \times 1$. The final Fully Connected (FC) layer transforms the features into binary class predictions, when Adam optimizer [25] was used for training.

#### *3.2.2. DEFORMISE-DenseStream*

The DEFORMISE-DenseStream model is a custom 3D CNN architecture (detailed in Fig. 1) that progressively increases channel depth (64 to 256) while maintaining aggressive pooling to capture multi-scale features. This design differs from standard progressive networks by maintaining dual channel layers initially, allowing for richer low-level feature extraction before expansion. Its architecture employs a sequence of convolutional, pooling, normalization, and fully connected layers to extract relevant features and facilitate classification tasks.

In the model's architecture, the convolutional segment of the network consists of 4 convolutional layers, each applying a 3D convolution operation to the input. A 3D convolution with 64 output channels, a kernel size of $3 \times 3 \times 3$, and padding set to 1 are used in the network's Convolutional Layer 1. A ReLU activation function is employed after this layer to add non-linearity. The max pooling operation with a kernel size of two completes the layer sequence. Similar to the first convolution, the second convolution uses a 3D convolution with the same kernel size and padding and keeps the number of output channels at 64. Max pooling, BN, and ReLU activation are also present in this layer. Progressing to Convolutional Layer 3, the network increases the output channels to 128, allowing for the construction of more intricate representations. The kernel size and padding remain consistent at 3×3×3 and 1, respectively. As with the previous layers, this one is succeeded by the ReLU activation, max pooling and BN, maintaining the structural integrity of the network's design. The Convolutional Layer 4 further escalates the complexity by increasing the number of output channels to 256. It upholds the kernel size of $3 \times 3 \times 3$ and padding of 1. The sequence of ReLU, max pooling and BNn follows.

Following the convolutional segment, the network employs an Adaptive Average Pooling layer, performing global average pooling to reduce each feature map to a single value. This layer transforms the 3D feature maps into a 1D array of 256 elements, setting the stage for the classification process. The network's FC section comprises two linear layers. The first expands the feature set from 256 to 512 units and includes a ReLU activation and a Dropout layer [26] with a rate of 0.3 to mitigate overfitting. Again, Adam optimizer was selected for training purposes.

### 3.2.3. DEFORMISE-CompactFocusNet

The DEFORMISE-CompactFocusNet model inherits the core principles of the EfficientNet design [27], as presented from (9) to (13), and a scalable architecture, but extends these concepts to handle 3D volumetric data effectively.

Step 1: Expansion

$$F_{\text{exp}} = \text{Swish}\left(\text{BatchNorm3d}\left(\text{Conv3d}_{1\times1\times1}(F_{\text{in}})\right)\right) \tag{9}$$

Step 2: Depthwise Convolution

$$F_{\text{dw}} = \text{Swish}\left(\text{BatchNorm3d}\left(\text{Conv3d}_{\text{depthwise}}(F_{\text{exp}})\right)\right) \tag{10}$$

Step 3: Squeeze-and-Excitation

$$F_{\text{se}} = \text{SEBlock3D}(F_{\text{dw}}) \tag{11}$$

Step 4: Projection

$$F_{\text{out}} = \text{BatchNorm3d}\left(\text{Conv3d}_{1\times1\times1}(F_{\text{se}})\right) \tag{12}$$

Step 5: Residual Connection

$$F_{\text{final}} = F_{\text{in}} + F_{\text{out}} \tag{13}$$

where $F_{\text{in}}$ represents the input feature map, $F_{\text{exp}}$, $F_{\text{dw}}$, $F_{\text{se}}$ and $F_{\text{out}}$ denote the feature maps at different stages within the block.

The model begins with a Stem Layer, consisting of a 3D convolutional layer with 32 output channels, a kernel size of 3, a stride of 2, and padding of 1. The stem also includes BN and the Swish activation function, a smooth, non-linear function that helps the model capture complex patterns, as shown in (14) below:

$$\text{Swish}(x) = x \cdot \sigma(x) \tag{14}$$

where $\sigma(x)$ is the sigmoid function, defined as $\sigma(x) = \frac{1}{1+e^{-x}}$. In this equation, $x$ is the input to the activation function. The sigmoid function outputs a value between 0 and 1, which scales the input $x$.

Following the stem, the model employs a series of blocks, as in Fig. 1. Each block typically consists of an expansion phase (using $1 \times 1 \times 1$ convolutions to increase the number of channels), a depthwise 3D convolution for spatial feature extraction, and a squeeze-and-excitation (SE) block that adaptively recalibrates channel-wise feature responses. The expansion ratio in these blocks dictates the degree of channel expansion in the first phase of the block. The depthwise convolutions, characterized by their kernel size and stride, are responsible for capturing spatial hierarchies in the data. The kernel sizes vary among the blocks, adapting the field of view to different spatial contexts.

The model's head section starts with a 3D convolution that expands the channels to 1280, followed by batch normalization and Swish activation. This is followed by a Flatten operation that maps the features to the desired number of classes for the classification task. As in previous models, Adam was selected as the optimizer.

### 3.2.4. Confidence-based Committee

A Confidence-based Committee approach is employed to leverage the predictive strengths of DEFORMISE-ResidualFlow, DEFORMISE-DenseStream and DEFORMISE-CompactFocusNet models. This committee operates by analyzing the predictions from each of these models for a given input and selecting the prediction with the highest confidence - "C" in (15). Confidence scores are computed to 5 decimal places precision from the softmax outputs, ensuring discrimination between models even with very similar confidence levels. As mentioned, confidence is derived from the softmax function applied to the models outputs, indicating the models certainty in their decisions.

$$\text{Pred} = \max(\text{C}_{\text{DEFORMISE-ResidualFlow}}, \text{C}_{\text{DEFORMISE-DenseStream}}, \text{C}_{\text{DEFORMISE-CompactFocusNet}}) \tag{15}$$

In the rare case where two models share identical maximum confidence (to 5 decimal places) but disagree on the prediction, the third model serves as the tie-breaker. In the extremely rare case where all three models have identical confidence scores but different predictions (both rare cases did not occur during validation), we apply majority voting where the prediction made by two models overrides the single dissenting model.

## 4. Experimental outcomes

### 4.1. Dataset description

Data from the OASIS1 [8] and OASIS2 [9] databases, which are both important sources in neuroimaging studies of brain aging and dementia, were used in the context of this investigation. The OASIS1 dataset, which includes cross-sectional MRI data from 416 participants between the ages of 18 and 96, is a publicly available collection of MRI data in the Neuroimaging Informatics Technology Initiative (NIfTI) format. This dataset includes a wide range of people, from those who are intellectually normal to those who have different degrees of cognitive impairment. Every participant in OASIS1 is represented by a single visit, during which four distinct 1.25 mm-thick T1-weighted MRI images are obtained and saved in NIfTI format. Complementing the OASIS1 dataset, OASIS2 provides longitudinal MRI data, also in NIfTI format, from 150 subjects aged between 60 and 96 years. Subjects in the OASIS2 dataset underwent MRI scans over two to five visits, with each visit spaced at least one year apart. Like in OASIS1, the MRI scans in OASIS2 also feature a slice thickness of 1.25 mm. This longitudinal approach, comprising 373 MRI sessions in total, is invaluable for studying the progression of neuro-degenerative diseases such as dementia.

Both datasets include vital metadata, featuring key patient characteristics such as age and Clinical Dementia Rating (CDR). The CDR scale categorizes subjects into classes 0, 0.5, 1, and 2, corresponding to no, very mild, mild, and moderate dementia, respectively. This classification provides a standardized measure of dementia severity, enabling the correlation of neuroimaging findings with the progression of cognitive impairment, greatly enriching the research's depth and applicability.

### 4.2. Evaluation metrics

In evaluating the performance of our classifier for distinguishing between demented and non-demented patients, we employed metrics derived from the confusion matrix. A confusion matrix offers a clear tabular representation of a classification algorithm's performance. In our specific binary classification context, the matrix components are True Positives (TP), True Negatives (TN), False Positives (FP), and False Negatives (FN). Here, TP denotes correctly identified demented cases, TN represents accurately identified non-demented cases, FP comprises non-demented cases mistakenly classified as demented, and FN includes demented cases incorrectly classified as non-demented.

Based on the outputs of this confusion matrix, we primarily focused on accuracy. Accuracy provides a straightforward measure of the classifier's overall correctness by combining the rates of TP and TN against the total number of cases, as shown in (16).

$$Accuracy = \frac{TP + TN}{TP + TN + FP + FN} \tag{16}$$

In addition to accuracy, we will also employ sensitivity and specificity to assess the model's performance. These metrics help in evaluating the effectiveness of the classifier in identifying demented patients (sensitivity) (17) and non-demented patients (specificity) (18) accurately.

$$\text{Sensitivity} = \frac{TP}{TP + FN} \tag{17}$$

$$\text{Specificity} = \frac{TN}{TN + FP} \tag{18}$$

**Table 1**
Evaluation & training time of models.

| Model | Accuracy ± std (%) | Sensitivity ± std (%) | Specificity ± std (%) | Training time (min) |
|---|---|---|---|---|
| DEFORMISE-ResidualFlow | 79.41 ± 1.86% | 74.31 ± 2.08% | 86.18 ± 4.70% | 622 |
| DEFORMISE-DenseStream | 85.29 ± 2.63% | 79.01 ± 2.73% | 93.49 ± 3.97% | 1360 |
| DEFORMISE-CompactFocusNet | 88.24 ± 4.16% | 86.00 ± 4.94% | 91.13 ± 5.58% | 499 |
| Committee | 94.12 ± 3.22% | 93.81 ± 3.73% | 94.50 ± 3.34% | |

**Table 2**
Comparative table.

| Method | Approach | Dataset | Accuracy |
|---|---|---|---|
| Saratxaga, 2021(BrainNet3D) [13] | 3D subject | OASIS-2 | 84.00% |
| Saratxaga, 2021(BrainNet3D) [13] | 3D subject | OASIS-1 | 84.00% |
| Dhinagar, 2021 [11] | 3D subject | OASIS-1 | 74.20% |
| Yagis, 2020 [12] | 3D subject | OASIS-1 | 69.90% |
| Wen, 2020 [14] | 3D subject | OASIS-1 | 68.00% |
| Saratxaga, 2021(ResNet18) [13] | 2D slice | OASIS-2 | 93.00% |
| Maqshood, 2019 [15] | 2D slice | OASIS-2 | 89.66% |
| Saratxaga, 2021(BrainNet2D) [13] | 2D slice | OASIS-1 | 84.00% |
| Saratxaga, 2021 (BrainNet2D) [13] | 2D slice | OASIS-2 | 83.00% |
| Wen, 2020 [14] | 2D slice | OASIS-1 | 68.00% |
| **Confidence-based Committee (Ours)** | **3D subject** | **OASIS-1 & OASIS-2** | **94.12%** |

**Table 3**
ADNI data selection criteria.

| Criterion | Description |
|---|---|
| Image types | Original |
| Study/Visit | ADNI Screening |
| Age (years) | ≥60 |
| Modality | MRI |
| Acquisition type | 3D |
| Weighting | T1 |
| Slice thickness (mm) | 1.2 |
| Protocol | MPRAGE |

**Table 4**
Comparative performance metrics of confidence-based committee on OASIS and ADNI datasets.

| Dataset | Accuracy (%) | Sensitivity (%) | Specificity (%) |
|---|---|---|---|
| OASIS-1 & OASIS-2 | 94.12% | 93.81% | 94.50% |
| ADNI | 90.96% | 95.73% | 80.73% |

### 4.3. Results

The training of the models was performed using a 5-fold cross-validation approach. For each fold, the model underwent training for 20 epochs, with a focus on monitoring the validation loss. The final evaluation of the models was conducted on a test set comprising 34 patients. Our confidence-based committee system aggregated the predictions from the individual models, taking into account their confidence levels, to arrive at a final diagnosis for each patient. Table 1 presents a concise summary of the accuracy, sensitivity and specificity results for each model, alongside the training times, enhanced with the standard deviation(std) for these metrics, providing insight into the variance observed across the folds.

The Confidence-based Committee, as described in the Table 1, effectively combines the results of the above models, leading to a notable increase in accuracy. While the individual models achieved average accuracy from 79.41% to 88.24%, the Committee model surpassed these with a remarkable 94.12%. Similar results are observed for sensitivity and specificity, where the Committee model achieves an average sensitivity of 93.81% and specificity of 94.50%. This underscores the ability of the to Committee model to identify both classes.

Table 2 provides a comparative view of the performance of various models developed for the binary classification of patients as demented or non-demented. The results in Table 2 are as reported in the original publications, each employing different preprocessing pipelines. It delineates not only the accuracy of each model but also the type of approach utilized - whether the model processes volumetric 3D data or 2D slices - as well as the specific dataset employed, be it OASIS-1 or OASIS-2. Our model, the Confidence-based Committee, employs a 3D Subject approach, integrating data from both OASIS-1 and OASIS-2 datasets. It has successfully outperformed all previous models tested solely on the OASIS datasets, achieving a superior accuracy of 94.12%. This holds true regardless of whether the previous models used 3D or 2D data for their classification tasks.

#### 4.3.1. Validation

To ensure the validity of our findings, we conducted a validation study using an external dataset, specifically the ADNI dataset [10]. The selection criteria for the data from the ADNI dataset were carefully curated to match the specific requirements of the current study and are detailed in Table 3.

Following the selection criteria of Table 3 we selected patients whose related metadata included the presence of Clinical Dementia Rating (CDR). To ensure the integrity of our analysis, we removed duplicate entries and follow-up examinations of the same patient. Finally, after categorizing our patients into classes based on their CDR scores — Demented (those with a CDR of 0.5 or 1) and Non-Demented (those with a CDR of 0) — we concluded with a total of 685 patients. We manipulated the Digital Imaging and Communications in Medicine (DICOM) data from the ADNI dataset, extracting the coronal slices per patient using pydicom python library. Then, we implemented the slice selection methodology described in the current study, Fig. 1, selecting the more suitable 140 continues slices, as shown in Fig. 4.

Finally, we proceeded with inference using the best-performing model of our Confidence-based Committee. As demonstrated in Table 4, our approach managed to achieve significant results in the ADNI subset as well, confirming the generalization capabilities of our methodology with an accuracy of 90.96%. It is also clear that while the percentage of correct predictions per class remains relatively stable, the unbalance in the dataset composition has influenced the higher sensitivity (95.73%) and lower specificity (80.73%) metrics. To statistically validate the committee's performance advantage, we evaluated each individual model on both the OASIS test set and ADNI validation set. Individual model accuracies were 79.41%, 85.29%, and 88.24% for DEFORMISE-ResidualFlow, DEFORMISE-DenseStream, and DEFORMISE-CompactFocusNet respectively on OASIS, and 80.89%, 81.17%, and 88.16% on ADNI. McNemar's test [28] comparing the Committee against DEFORMISE-CompactFocusNet across both datasets ($n = 719$ combined) yielded $\chi^2 = 21.0, p < 0.001$, demonstrating statistically significant improvement with the Committee correctly classifying 21 additional patients.

Furthermore, we conducted an analysis of our Preprocessing Pipeline by comparing our novel approach for the pre-selection of MRI slices against a baseline technique. This baseline involves selecting 140 contiguous MRI slices based on the entropy criterion using (19):

$$H = -\sum_{i=1}^{n} p_i \log_2(p_i) \tag{19}$$

where $p_i$ represents the probability of occurrence of the $i$th intensity value in the image, and $n$ is the number of distinct pixel intensities.

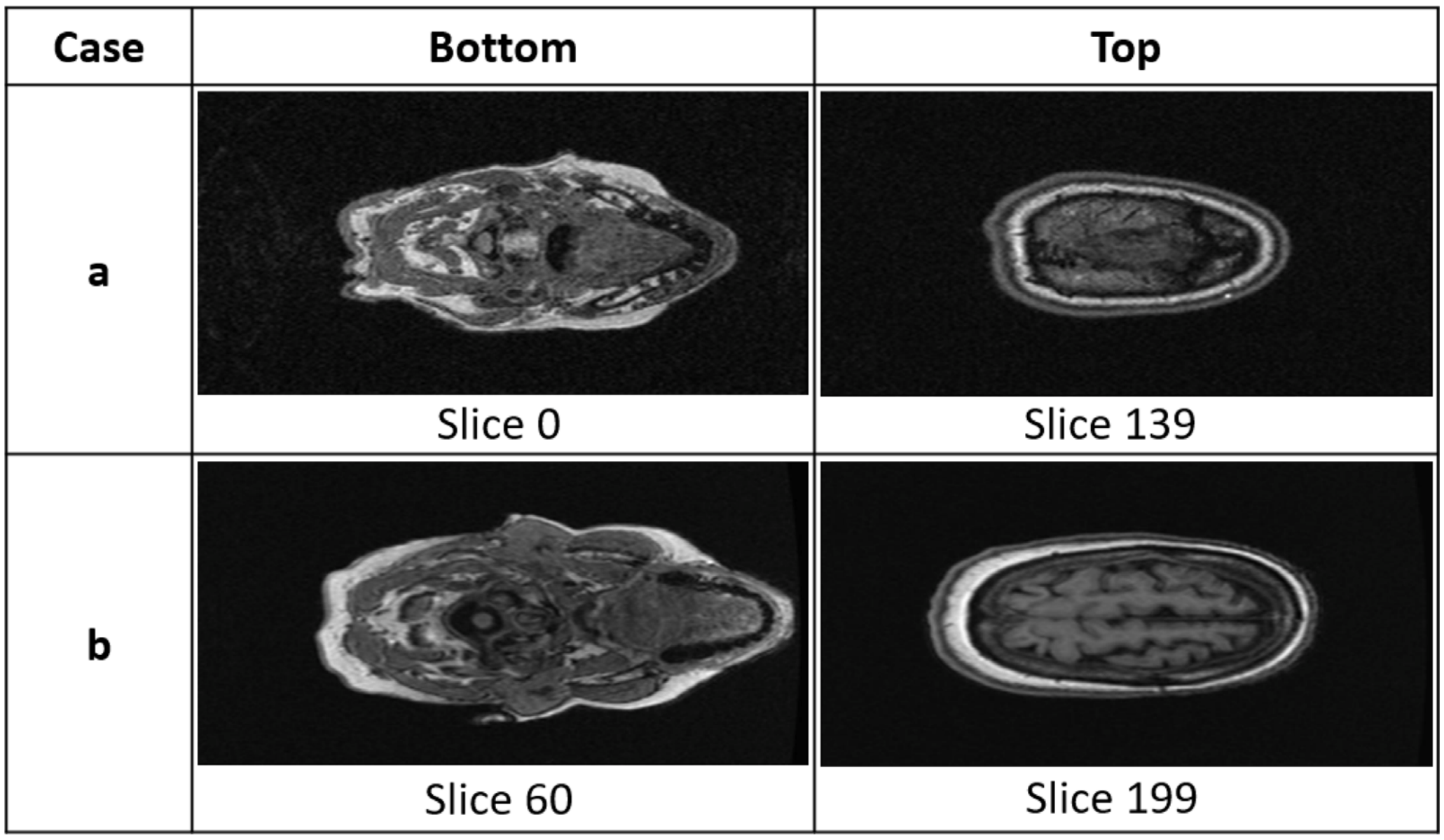

**Fig. 4.** Indicative examples of slice selection of ADNI patients: (a) patient 099_S_0470 classified as Demented and (b) patient 136_S_0086 classified as Non Demented.

Manual inspection revealed that the entropy-based slice selection frequently included non-diagnostic regions. As demonstrated in Fig. 5, the entropy method often selects slices containing skull, neck tissue, partial brain coverage or empty slices while our edge-based method consistently captures the complete brain volume with relevant diagnostic structures. This visual comparison validates that edge density is a superior proxy for identifying diagnostically relevant slices compared to entropy alone.

Additionally we analyzed the individual model contributions within our Confidence-based Committee, consisting of DEFORMISE-Residual Flow, DEFORMISE-DenseStream and DEFORMISE-CompactFocusNet for the classification of brain MRIs into demented and non-demented categories. This examination aligns perfectly with our previously reported accuracy. The DEFORMISE-ResidualFlow, with an average accuracy of 79.41%, contributes 3 predictions to the committee's decisions. The DEFORMISE-DenseStream, demonstrating a higher average accuracy of 85.29%, is responsible for 6 predictions. Most significantly, the DEFORMISE-CompactFocusNet, with the highest individual average accuracy of 88.24%, contributes to 25 predictions. This proportional representation of each model's predictions to their accuracy levels validates our initial findings and highlights the effectiveness of our committee-based approach.

As final validation step we have incorporated explainable AI(XAI) techniques, specifically focusing on visualizing areas of attention in the model predictions. This aspect of our validation work centered on the model within our committee that demonstrated the highest confidence for a specific prediction, in this case, the DEFORMISE-CompactFocusNet model. We applied this approach to a correct prediction made by the model for patient 0022 from the OASIS1 dataset. The core of our XAI approach involved implementing the Gradient weighted Class Activation Mapping (Grad-CAM) technique [29]. This method enabled us to gain insights into which specific parts of the MRI scans were pivotal in the model's decision-making process. By attaching hooks to the model's target layer, we captured the necessary activation and gradients during the forward and backward passes of the model. Following this, we generated a heatmap from these activation, highlighting the areas of the brain scan that were most influential in the model's predictions.

Based on the findings of Ferreira et al. [30], brain atrophy in specific regions, such as the hippocampus, entorhinal cortex, cingulate gyrus and other areas of the cerebral cortex, is crucial for the detection of dementia. Recent studies have particularly emphasized ventricular enlargement as a robust early biomarker. Lee et al. [31] demonstrated that initial ventricular size significantly predicts the rate of cognitive decline in Alzheimer's disease patients, with larger baseline ventricles associated with faster disease progression. In the same figure, we also include the attention map generated from our predictive model. This map distinctly marks the areas of the brain that our model focuses on when making a prediction. The attention map aligns closely with the ventricles of the brain that are known to undergo atrophy in dementia as demonstrated in Fig. 6. This alignment validates our model's accuracy.

#### 4.3.2. *Ablation study*

For our ablation study, we will proceed along two axes. The first axis concerns our Preprocessing Pipeline and the methodology for selecting the optimal 140 slices. This aspect of the study is crucial for understanding the efficacy and precision of our technique in handling brain MRI scans. In our approach, based on a thorough review of relevant literature, we have identified that 140 slices, or 17.5 cm, is the ideal depth for brain MRIs slices. This depth is critical to ensure that all necessary brain areas are included in the scans for accurate diagnostics and analysis. It is important to note that this selection technique can be applied to all datasets extracted using similar protocol methods as those used in the OASIS 1, OASIS 2 and ADNI which produce slices of 1.25 mm with no inter-slice gap. The MRI protocol employed in these studies, known as Magnetization Prepared Rapid Gradient Echo (MP-RAGE) [32], is crucial for achieving high-resolution images that facilitate detailed anatomical analyses.

To validate this choice, our study includes a case analysis as illustrated in Fig. 7. This analysis, which is primarily qualitative, demonstrates the implications of varying the number of slices in an MRI scan. When we reduce the number of slices to 120, there appears to be a compromise in the MRI scan's comprehensiveness. Notably, the uppermost slice in the 120-slice shows a reduction in the brain area coverage. This observation suggests that decreasing the number of slices might lead to the exclusion of sections of the brain, potentially affecting the completeness of the diagnostic information. Conversely, increasing the slice count to 160 introduces its own set of challenges. In this scenario, the scan begins to include unnecessary elements. The bottom slice becomes blurry, diminishing the overall clarity of the scan, and the top part of the scan includes excessive bone structure that is not pertinent to the brain areas of interest. This not only adds to the data volume but also potentially distracts from the crucial brain areas that need to be analyzed. Thus, our study underscores that the selection of 140 slices is not arbitrary but a carefully balanced choice. This volume ensures the inclusion of all necessary brain areas while avoiding the inclusion of irrelevant or unclear portions, as also visible in Fig. 7.

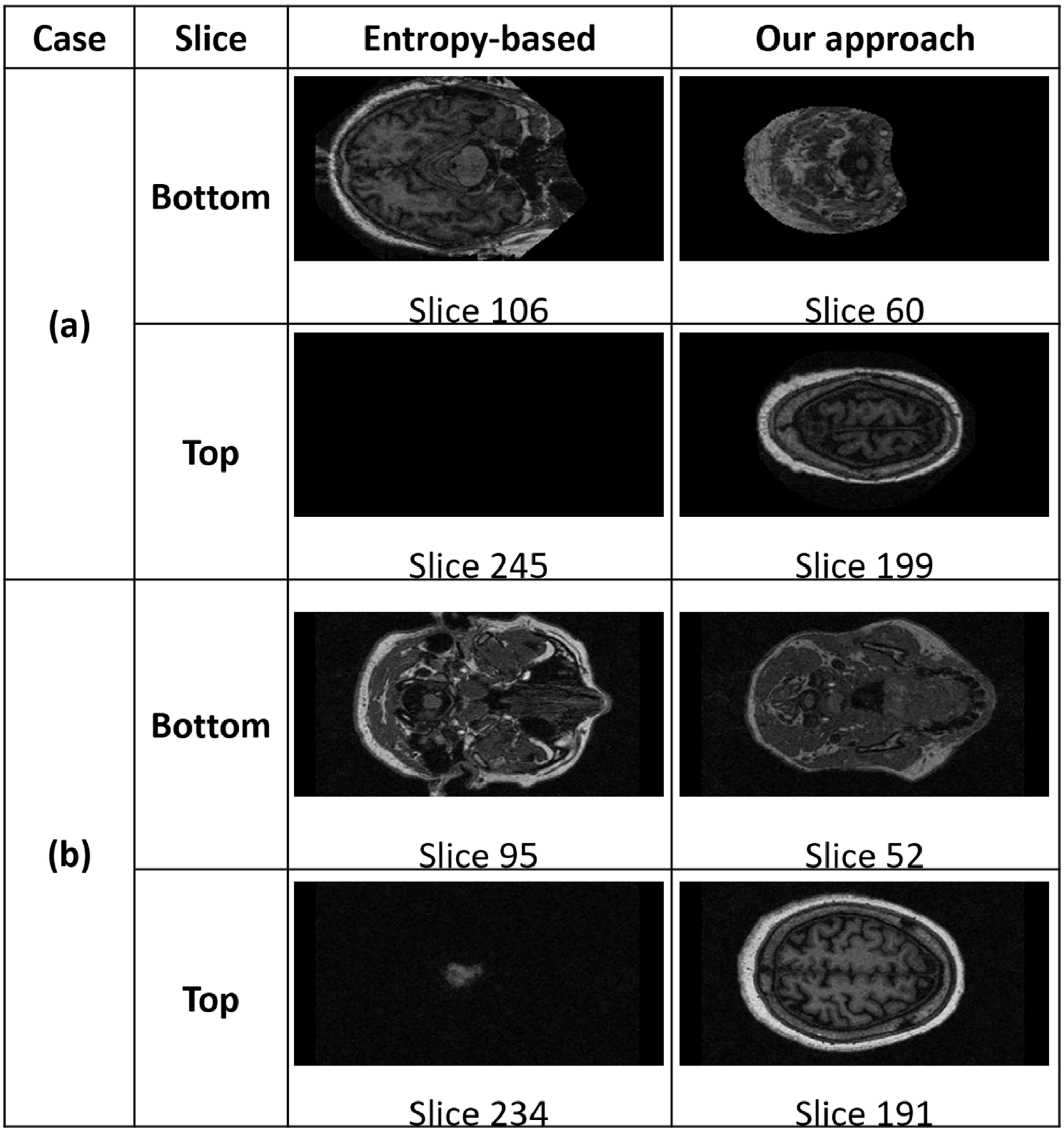

**Fig. 5.** Comparative examples of slice selection: Entropy-Based Method vs. Our approach for (a) patient 0223 from the OASIS1 dataset and (b) patient 0159 from the OASIS2 dataset.

**Table 5**
Quantitative performance comparison for different slice counts.

| Slice count | Accuracy (%) | Sensitivity (%) | Specificity (%) | p-value (paired t-test, 140 vs. slice count) |
|---|---|---|---|---|
| 120 | 87.65 | 84.38 | 91.67 | 0.045 |
| 140 | 94.12 | 93.81 | 94.50 | – |
| 160 | 91.18 | 90.63 | 91.89 | 0.26 |

**Table 6**
Per-patient inference analysis: 140-slice vs. 256-slice.

| Model | 140 slices (s) | 256 slices (s) | Speedup | FLOPs (GFLOPs) |
|---|---|---|---|---|
| DEFORMISE-ResidualFlow | 9.84 | 11.34 | 1.15× | 380/720 |
| DEFORMISE-DenseStream | 10.08 | 11.48 | 1.14× | 550/1000 |
| DEFORMISE-CompactFocusNet | 9.28 | 10.86 | 1.17× | 180/340 |
| Committee aggregation | 0.03 s (negligible) | | | – |

Table 5 demonstrates that 140 slices achieves optimal performance, with accuracy declining for both 120 slices (incomplete brain coverage) and 160 slices (inclusion of non-diagnostic regions). The performance degradation confirms our anatomically-driven slice selection. To perform this, we modified our preprocessing pipeline to select 120 and 160 continuous slices using the same edge-based methodology. Paired t-tests using 5-fold cross-validation results confirmed that 140 slices significantly outperforms 120 slices ($t(4) = 2.86$, $p = 0.045$), validating our anatomically-informed selection. The difference between 140 and 160 slices did not reach statistical significance ($t(4) = 1.30$, $p = 0.26$), though computational efficiency advantages (Table 6) and qualitative analysis (Fig. 7) support 140 slices as optimal.

Beyond diagnostic accuracy, we also quantified the computational efficiency (per-patinet inference) of our 140-slice selection. Table 6 presents measurements conducted on our test infrastructure (8× NVIDIA A16 GPUs with 16 GB VRAM each, 64 GB system RAM), comparing complete pipeline performance (preprocessing + inference) for 140-slice versus 256-slice configurations. Note that 256-slice models required architectural modifications to accommodate the larger input volume.

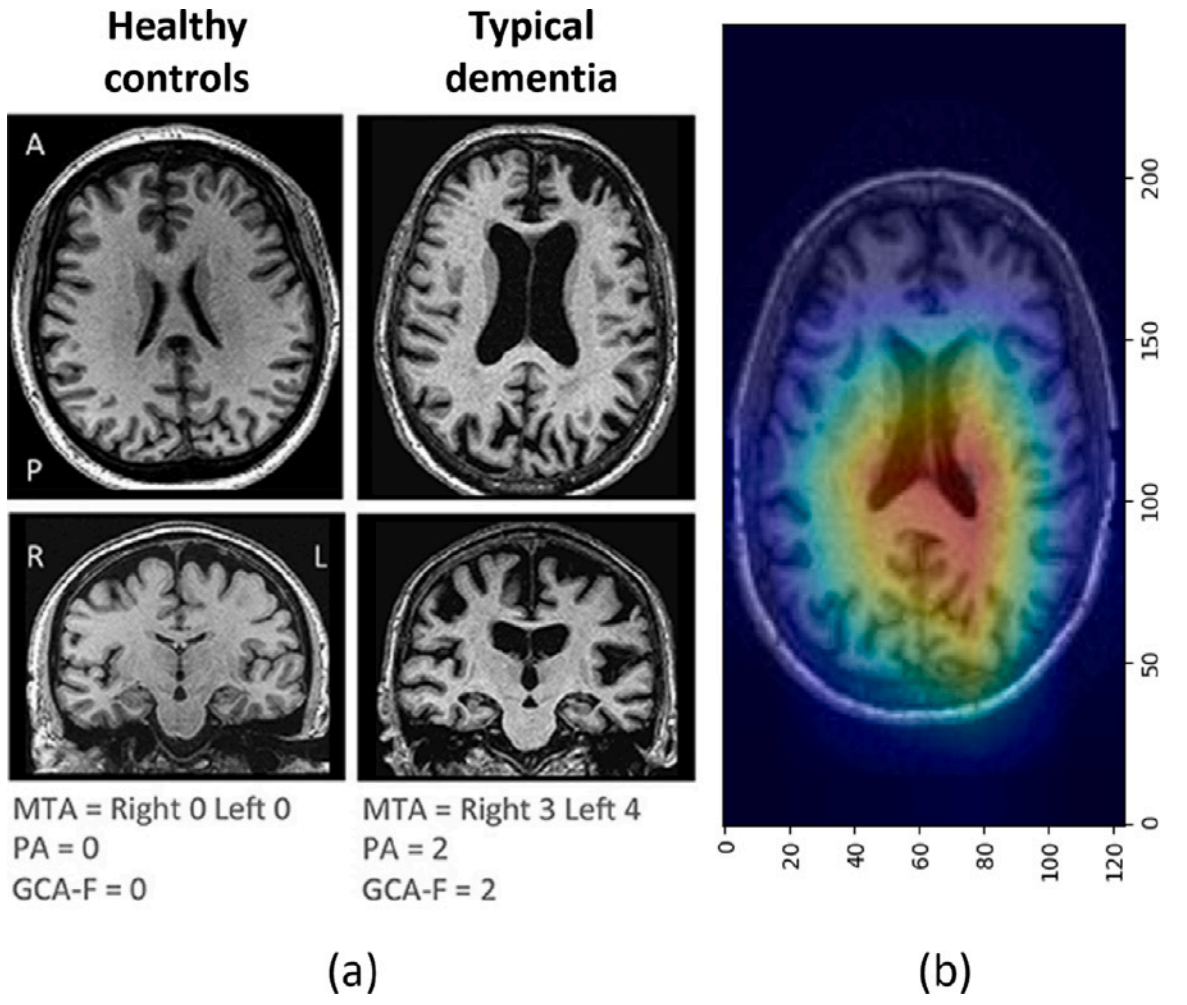

**Fig. 6.** (a) Comparison between normal brain and dementia case [30], (b) Attention map from the predictive model, with red areas indicating key regions for classifying a patient as demented.

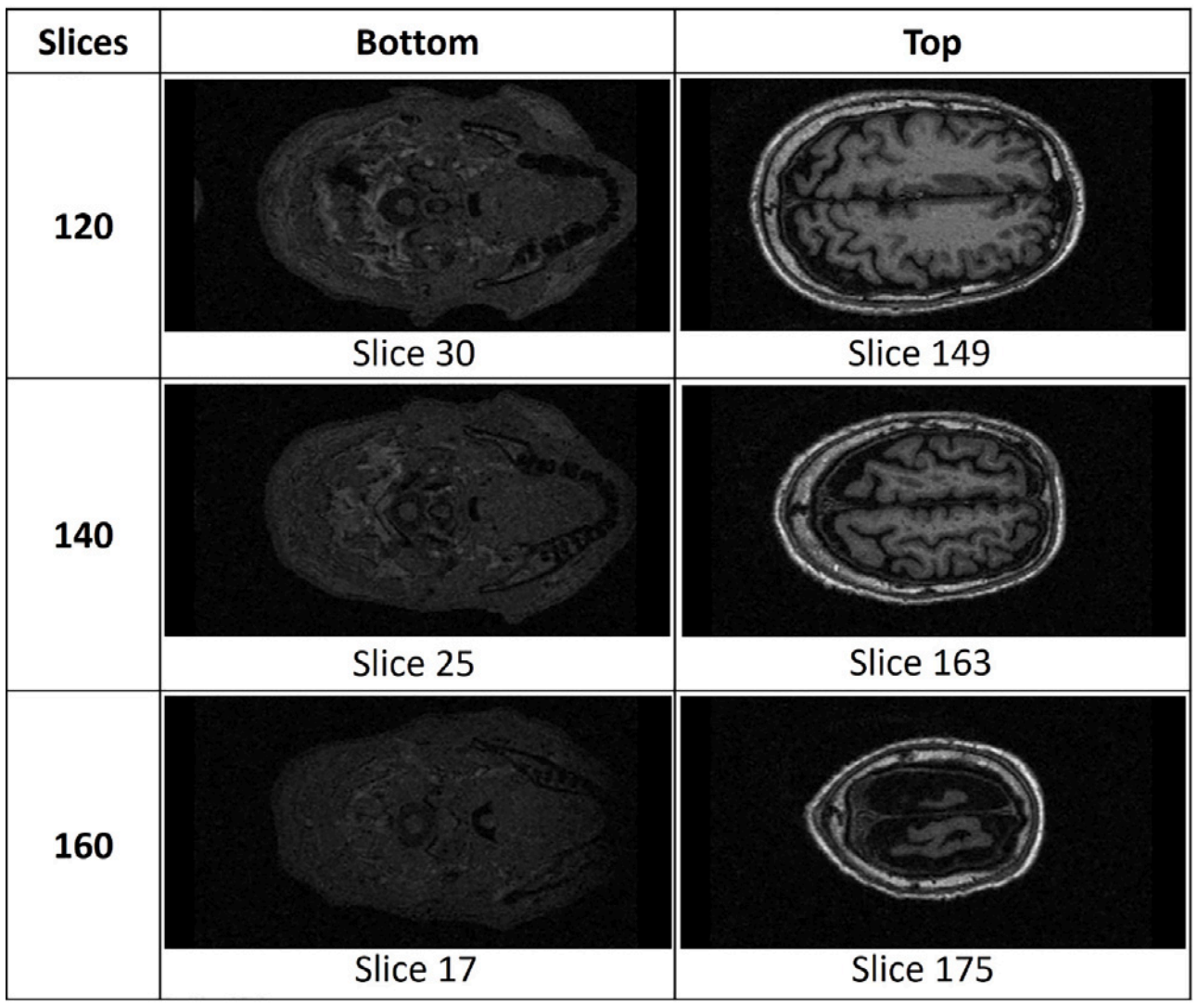

**Fig. 7.** Comparative display of top and bottom slices in brain MRIs with varying slice counts based on Algorithm 1 of patient 0009 from the OASIS2 dataset.

Results show consistent efficiency gains of 1.14–1.17× (14%–17% faster processing) across all models. FLOPs are reduced by approximately 47%–50%, and GPU memory requirements decrease from 5.5–6.5 GB to 2.5–3.5 GB. Committee aggregation overhead remains negligible (0.03 s).

For the second axis of ablation study, the focus shifts to evaluating the performance of a confidence-based committee comprised of three distinct models in the final decision-making process. This part of the study aims to demonstrate the synergy and enhanced performance achieved through the integration of these three models compared to any pair-wise combination. At this stage, we will not only present the individual average accuracy of each model but building on this, the ablation study will explore how different pairings of these models perform relative to the collective operation of all three.

Table 7 presents the performance of individual models, confidence-based pair-wise combinations and the tri-model committee in terms of average accuracy. Individually, each model demonstrates different levels of accuracy, DEFORMISE - CompactFocusNet model leading at 88.24%, followed by the DEFORMISE - DenseStream model at 85.29%, and the DEFORMISE - ResidualFlow model at 79.41%.

**Table 7**
Average accuracy comparison of individual and combined Dem3D models.

| DEFORMISE-ResidualFlow | DEFORMISE-DenseStream | DEFORMISE-CompactFocusNet | Average accuracy |
|---|---|---|---|
| ✓ | | | 79.41% |
| | ✓ | | 85.29% |
| | | ✓ | 88.24% |
| ✓ | ✓ | | 82.35% |
| ✓ | | ✓ | 88.24% |
| | ✓ | ✓ | 91.18% |
| ✓ | ✓ | ✓ | 94.12% |

When examining the pair-wise confidence-based committee combinations, an improvement in accuracy is observed. The combination of DEFORMISE - DenseStream and DEFORMISE - CompactFocusNet models achieves the highest accuracy among pairs, reaching 91.18%. This is followed closely by the combination of DEFORMISE - ResidualFlow and DEFORMISE - CompactFocusNet at 88.24% and then by the combination of DEFORMISE - ResidualFlow and DEFORMISE - DenseStream at 82.35%. These results indicate a synergistic effect, where the combination of models compensates for individual weaknesses and enhances overall performance.

Most crucially, the tri-model committee, integrating all three models — DEFORMISE - ResidualFlow, DEFORMISE - DenseStream, DEFORMISE - CompactFocusNet achieves the highest average accuracy of 94.12%. This outcome validates that the collective decision-making of the three models outperforms any individual or pair-wise model combination. The superiority of the tri-model configuration is attributed to the diverse strengths and analytical perspectives each model contributes, leading to a more comprehensive and accurate decision-making process.

## 5. Conclusion

This study successfully introduced DEFORMISE, a novel methodology for the binary classification of demented and non-demented patients using 3D brain MRI scans, achieving a notable milestone with an average accuracy of 94.12% on the combination of OASIS1 and OASIS2 datasets. The heart of this achievement lies in the preprocessing technique along with the confidence-based classification committee, a harmonious integration of three distinct models — DEFORMISE - CompactFocusNet, DEFORMISE - DenseStream and DEFORMISE - ResidualFlow. This methodology not only enhanced the accuracy of dementia diagnosis but also notably reduced the computational load, thereby marking a step forward in the practical application of AI in clinical settings. Future work could explore comparative analyses between our slice selection and established brain segmentation approaches, quantifying the trade-offs between computational efficiency and potential performance gains. Additionally, integrating 2D slice-level methods with our 3D volumetric models in a hybrid committee could potentially leverage complementary error modes, as 2D and 3D approaches may capture different diagnostic features. The approach of selectively processing MRI scans to focus on the most pertinent data further underscores the innovative angle of this research.

## CRediT authorship contribution statement

**Nikolaos Ntampakis:** Writing – original draft, Software, Resources, Methodology, Investigation, Data curation, Conceptualization. **Konstantinos Diamantaras:** Writing – review & editing, Supervision. **Ioanna Chouvarda:** Writing – review & editing, Supervision. **Vasileios Argyriou:** Writing – review & editing, Supervision. **Panagiotis Sarigiannidis:** Writing – review & editing, Supervision.

## Ethical statement

All data used in this study were obtained from publicly available, de-identified databases. The OASIS data collection was approved by the Washington University School of Medicine Human Studies Committee, and written informed consent was obtained from all participants [8,9]. The ADNI study was approved by the Institutional Review Boards of all participating institutions, and written informed consent was obtained from all participants according to the Declaration of Helsinki (ADNI protocol available at adni.loni.usc.edu) [10]. As this study involved only secondary analysis of publicly available, de-identified data, no additional ethical approval was required in accordance with institutional and national research committee guidelines.

## Declaration of competing interest

The authors declare the following financial interests/personal relationships which may be considered as potential competing interests: Nikolaos Ntampakis reports financial support was provided by Horizon Europe. If there are other authors, they declare that they have no known competing financial interests or personal relationships that could have appeared to influence the work reported in this paper.

## Acknowledgments

This project has received funding from the European Union's Horizon Europe research and innovation programme under grant agreement No. 101095435 (REALM).

## Data availability

The data have been published as open data.